\documentclass[lettersize,journal]{IEEEtran}
\usepackage{amsmath,amsfonts}
\usepackage{algorithmic}
\usepackage{algorithm}
\usepackage{array}
\usepackage[caption=false,font=normalsize,labelfont=sf,textfont=sf]{subfig}
\usepackage{textcomp}
\usepackage{stfloats}
\usepackage{url}
\usepackage{verbatim}
\usepackage{graphicx}
\usepackage{cite}
\usepackage{nomencl}
\usepackage{tabularx, booktabs}
\usepackage{multirow}
\usepackage{amsthm}
\usepackage{pifont}
\newcolumntype{C}{>{\centering\arraybackslash}X} 
\makenomenclature

\newtheorem{thm}{Claim}[section]

\begin{document}

\title{Feeder Microgrid Management on an Active Distribution System during a Severe Outage}

\author{Valliappan Muthukaruppan,~\IEEEmembership{Student Member,~IEEE,} Ashwin Shirsat,~\IEEEmembership{Student Member,~IEEE,} 

Rongxing Hu,~\IEEEmembership{Student Member,~IEEE,} Victor Paduani,~\IEEEmembership{Student Member,~IEEE,} Bei Xu,~\IEEEmembership{Student Member,~IEEE,} 

Yiyan Li,~\IEEEmembership{Member,~IEEE,} Mesut Baran,~\IEEEmembership{Fellow,~IEEE,} Ning Lu,~\IEEEmembership{Fellow,~IEEE,} 

David Lubkeman,~\IEEEmembership{Fellow,~IEEE,} and Wenyuan Tang,~\IEEEmembership{Member,~IEEE,}
\thanks{\textit{Corresponding author}: Mesut Baran (\it{baran@ncsu.edu})}
\thanks{The authors are with Department of Electrical and Computer Engineering, 
North Carolina State University, Raleigh, NC 27695 USA (email: {vmuthuk2, ashirsa, rhu5, vdaldeg, bxu8, yli257, nlu2, dllubkem,
wtang8, baran}@ncsu.edu). This material is based upon work supported by U.S.
Department of Energy’s Office of Energy Efficiency and Renewable Energy (EERE) under Solar Energy Technologies Office Award Number DE-EE0008770.}}



\maketitle

\begin{abstract}
Forming a microgrid on a distribution system with large scale outage after a severe weather event is emerging as a viable solution to improve resiliency at the distribution level. This option becomes more attractive when the distribution system has high levels of distributed PV. The management of such feeder-level microgrid has however many challenges, such as limited resources that can be deployed on the feeder quickly, and the limited real-time monitoring and control on the distribution system. Effective use of the distributed PV is also challenging as they are not monitored and controlled. To handle these challenges, the paper proposes a 2-stage hierarchical energy management scheme to securely operate these feeder level micorgrids. The first stage of the scheme solves a  sequential rolling optimization problem to optimally schedule the main resources (such as a mobile diesel generator and battery storage unit). The second stage adopts a dispatching scheme for the main resources to adjust the stage-1 set-points closer to real-time. The proposed scheme has unique features to assure that the scheme is robust under highly varying operating conditions with limited system observability: (i) an innovative PV forecast error adjustment and a dynamic reserve adjustment scheme to handle the extreme uncertainty on PV power output, and (ii) an intelligent fuel management scheme to assure that the resources are utilized optimally over the multiple days of the restoration period. The proposed algorithm is tested on sample system with real-time data. The results show that the proposed scheme performs well in maximizing service to loads by effective use of all the resources and by properly taking into account the challenging operating conditions.
\end{abstract}


\section*{Nomenclature}
\addcontentsline{toc}{section}{Nomenclature}

\addcontentsline{toc}{subsection}{Parameters}
\begin{IEEEdescription}[\IEEEusemathlabelsep\IEEEsetlabelwidth{$V_1,V_2,V_3,V_4,V_5$}]
\item[$\overline{E}^{ES}_i$] kWh rating of ES $i\in\mathcal{N}^{ES}$
\item[$\left\{\underline{F}, \overline{F}\right\}_i$] Min/Max fuel limits of DG $i\in\mathcal{N}^{DG}$
\item[$\overline{S}^{ES/DG}_i$] kVA rating of ES/DG $i\in\mathcal{N}^{ES/DG}$
\item[$\left\{\underline{SoC}, \overline{SoC}\right\}_i$] Min/Max SoC limits of ES $i\in\mathcal{N}^{ES}$
\item[$\alpha_i, \beta_i$] Fuel consumption coefficients of DG
\item[$\gamma_t$] Reserve factor for GFM ES unit
\item[$\alpha^{MSD}$] Minimum service duration of load groups
\item[$\alpha^{up}$] Minimum up time for DG
\item[$w_i$] Priority weight for load groups
\item[$\theta_i$] Power factor angle for DG $i\in\mathcal{N}^{DG}$
\item[$C^{up}$] Start cost for DG
\item[$\lambda_n$] Forecast error normalization factor
\item[$F_f$] Final desired fuel reserve for DG

\end{IEEEdescription}

\addcontentsline{toc}{subsection}{Stage-1 Decision Variables}
\begin{IEEEdescription}[\IEEEusemathlabelsep\IEEEsetlabelwidth{$V_1,V_2,V_3,V_4,V_5$}]
\item[$\{P_{i,t,\phi}, Q_{i,t,\phi}\}^{D}$] Real and Reactive power load demand 
\item[$\{P_{i,t,\phi}, Q_{i,t,\phi}\}^{ES}$] Real and Reactive power output of ES 
\item[$\{P_{i,t}, Q_{i,t}\}^{DG}$] Real and Reactive power output of DG
\item[$P_{i,t,\phi}^{PV}$] Real Power output of BTM PV
\item[$x_{n,t}$] Status of switch connecting $n^{th}$ LG
\item[$y_{i,t}$] Status of switch controlling $i^{th}$ DG
\item[$C_{i,t}$] Start up cost of $i^{th}$ DG
\end{IEEEdescription}

\addcontentsline{toc}{subsection}{Stage-2 Decision Variables}
\begin{IEEEdescription}[\IEEEusemathlabelsep\IEEEsetlabelwidth{$V_1,V_2,V_3,V_4,V_5$}]
\item[$\{P_{i,k,\phi}, Q_{i,k,\phi}\}^{D}$] Real and Reactive power load demand 
\item[$\{P_{i,k,\phi}, Q_{i,k,\phi}\}^{ES}$] Real and Reactive power output of ES 
\item[$\{P_{i,k}, Q_{i,k}\}^{DG}$] Real and Reactive power output of DG
\item[$P_{i,k,\phi}^{PV}$] Real Power output of BTM PV
\item[$x_{n,k}$] Status of switch connecting $n^{th}$ LG
\item[$\hat{x}_{n,t}$] Stage-1 schedule of $n^{th}$ switch in $\Delta t$
\item[$\hat{P}_{i,t}^{DG}$] Stage-1 DG schedule in $\Delta t$
\item[$\hat{\epsilon}_t$] Forecast error correction factor in $\Delta t$
\end{IEEEdescription}

\section{Introduction}
\IEEEPARstart{R}{ecent} increase in extreme weather events encourage utilities to take measures to increase the resiliency especially at the distribution level in order to provide service during such extreme events \cite{preston2016}. One promising technology is the microgrids as it facilitates the operation of the healthy part of the system during an extended outage caused by an extreme event \cite{hussain2019, hirsch2018}. Formation of a microgrid becomes more attractive especially on a distribution system with high penetration of renewables, mainly distributed Photovoltaic (PV) \cite{booth2017}.

Another promising technology for forming a microgrid on a distribution feeder on demand is the mobile energy storage (MES) device as it provides the flexibility to change location depending on changing grid or customer conditions \cite{DOER2020, abdeltawab2019, Kim_2019}. Hence, it becomes quite feasible to form a feeder level microgrid during extended outage (day or more) \cite{Che_2019, Lei2018, Lei2019}. If the feeder has large penetration of distributed PV, then forming such a microgrid becomes more attractive \cite{DOER2020}.

Forming a microgrid on a feeder has unique challenges \cite{koutsoukis2019, choi2022}. The main feature is the mobile generation resources that can be brought to the location. A typical approach is to bring a MES with a diesel generator (DG). Both of these resources will have limited capacity and energy that needs to be carefully managed \cite{dugan2021}. Another issue is that the current distribution feeders have only limited circuit switches for control, which makes it challenging to ration the load when needed. Having a large amount of PV on the feeder can help considerably with the microgrid operation. The challenge with utilizing these resources is that most of these resources are not visible (i.e., not monitored) and they are quite intermittent. These conditions make it challenging to operate a feeder-level microgrid during an extended multi-day outage. This paper focuses on this problem.

Existing literature address only some of the challenges identified. In \cite{chen2016a, meng2020} authors focus on black start of a feeder-level microgrid by picking up outage load with limited resources and with cold load pickup consideration in \cite{chen2017, yuan2016}. In \cite{momen2021, poudel2019} load restoration over limited duration is considered and it uses direct load control to selectively restore critical loads. In \cite{kleinberg2011} a similar problem is considered with direct load control but behind-the-meter (BTM) PV is not included.

In \cite{liu2021, liu2021b} load restoration problem over limited duration is considered and a multi-microgrid formation and reconfiguration is adopted by leveraging Distributed Energy Resource (DER) flexibility. The DER flexibility is evaluated and implemented using an aggregator which coordinates with upstream distribution system operator for restoration. For realizing the flexibility of DERs and loads, author assumes energy storage (ES) devices collocated with the PV, thermostatically controllable loads like HVAC can be fully leveraged by the Distribution System Operator (DSO) at each house level. Such coordination, flexibility on demand side and concept of DER aggregators is yet to be implemented in many utilities. This framework also requires significant communication between the aggregators and the DSO. In \cite{koutsoukis2019, choi2022} realistic existing feeder infrastructure is considered in restoration but they do not include the DER uncertainty in the restoration process.

This paper aims at development of a comprehensive microgrid management scheme for the emerging case of operating a feeder level microgrid on an active distribution system to provide service during a severe outage. The main contribution of the proposed method is that all the main issues/challenges associated with such a microgrid is considered and a robust management scheme is developed to address these issues. The contributions include:

\begin{itemize}
  \item Realistic distribution system operating conditions are considered: limited load and PV visibility and controllability, and limited controllable switches on the feeder.
  \item To make best use of the main resources for the microgrid - a DG, and a MES - a novel DG fuel management scheme is adopted to ensure service to loads during peak load conditions over multiple days.
  \item To address the main challenge associated with estimating BTM PV variability, a new forecast error estimation and adjustment strategy to correct the high forecast error has been introduced which significantly increases the amount of PV utilization during restoration.
  \item A dynamic reserve adjustment strategy for the MES is also introduced in order to securely operated the microgrid under severe cloud cover events and to minimize the unintentional load shedding events.
\end{itemize}

The rest of the paper is organized as follows: sec-\ref{sec:probform} introduces the feeder microgrid management problem and the proposed scheme, sec-\ref{sec:MicrogridSecurity} outlines the new robustness enhancement strategies proposed to handle the issues associated with the BTM PV and multi-day restoration using limited resources. Section-\ref{sec:casestudy} illustrates the performance of proposed scheme with a case study which includes the IEEE 123 node system and operating conditions based on field data.

\section{Feeder Microgrid Management Scheme}\label{sec:probform}
\subsection{Operating Conditions}\label{subsec:OprCond}
To ensure realistic operating conditions for feeder level microgrid are included in the management scheme, the following conditions are considered:
\begin{itemize}
  \item Utility owned mobile devices MES and DG are quickly deployed at a proper location on the distribution feeder which has lost power. The MES and DG  are the main resources and are designed to be connected to feeder at substation or other proper location with necessary infrastructure. The microgrid controller with the energy management scheme can monitor and control these resources.
  \item The feeder has high level of BTM PV which will provide supplemental power for the microgrid. These resources are not visible to the microgrid controller, as BTM PV is typically not monitored. Hence, there is a need to forecast at least the net load (actual load minus the power from PV) for the management of the microgrid.
  \item The distribution feeder has only a limited remotely controllable sectionalizing switches. The microgrid controller can use these switches to adjust the load on the feeder, as these switches divide the loads in to load groups which can be disconnected as needed. The feeder has also some critical loads and they should have higher priority in the restoration.
\end{itemize}

\subsection{Energy Management Scheme}\label{subsec:EMS}
\begin{figure}[htpb]
  \centering
  \includegraphics[width=3.5in]{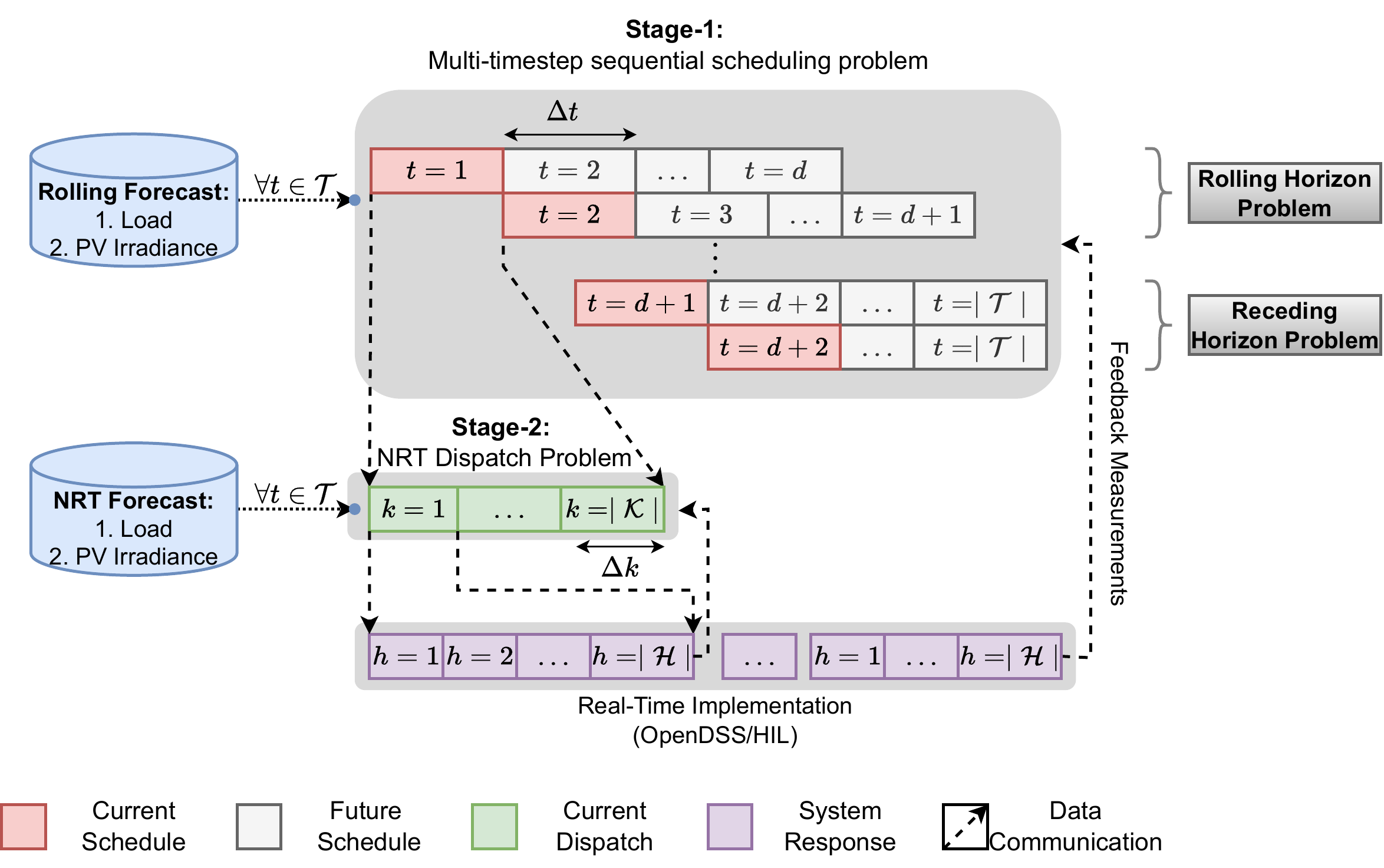}
  \caption{Proposed Energy Management Framework}
  \label{fig:EMSframework}\vspace{-0.6cm}
\end{figure}

We propose a 2-stage hierarchical energy management framework as presented in figure-\ref{fig:EMSframework}. Stage-1 is the scheduler to schedule the microgrid resources for the next period (such as half or one hour) by taking in to account future load and PV forecast. Stage-2 is the short term dispatching stage which determines the proper dispatch levels for MES and the DG for the next dispatch period (of 1 to 5 minutes). After every dispatch cycle ($\Delta k$) and scheduling cycle ($\Delta t$) the real-time (RT) measurements are used to update the initial condition of next dispatch or rolling horizon cycle respectively. These two stages are outlined below.

\subsubsection{Stage-1: Scheduling Problem}\label{subsec:stage1}
The stage-1 problem which solves a receding horizon optimization problem on last day of restoration and a rolling horizon problem on other days of restoration \cite{glomb2022}, to optimally allocate the available resources and control the amount of load to be picked up in each interval through load groups over the considered horizon which is typically 24 hours to ensure the adequate availability of resources over the entire restoration period. The important decision variables in this stage are the switch status $x_{n,t}$ and the scheduled DG output $P^{DG}_{i,t}$. The Load and PV at each node given by $P^{D/PV}_{i,t,\phi}$ is the stochastic variable obtained from forecast information. The objective function is shown in (\ref{eqn:obj}) where the first term maximizes the total expected load  groups to be served denoted by $x_{n,t}$ with higher priority to load groups with critical loads (which have higher weights $w_n$) and second term minimizes the start up cost of diesel generator.
\begin{equation}\label{eqn:obj}
   \max_{x} \quad \sum_{t\in\mathcal{T}} \left[ \sum_{i\in\mathcal{N}_n} x_{n,t} w_{n} \sum_{\phi\in\Phi}P_{i,t,\phi}^{D}\Delta t - \sum_{i\in\mathcal{N}^{DG}} C_{i,t} \right]
\end{equation}

The constraints are defined for $t\in\mathcal{T}$, $\phi \in \Phi$, and $n\in\mathcal{N}^{LG}$ unless explicitly stated. Equation (\ref{eqn:Pbalance}) highlights the real power balance in the network over entire time horizon $\mathcal{T}$. The  circuit switches  $x_{n,t}$ are the binary decision variables. The reactive power balance in the network is shows in (\ref{eqn:Qbalance}). It is assumed that all BTM PVs operate at unity power factor with no reactive power injection.
\begin{subequations}\label{eqn:PowerFlow}
  \begin{align}
  \sum_{i\in\mathcal{N}^{ES}} P_{i,t,\phi}^{ES} + \sum_{i\in\mathcal{N}^{DG}} \frac{P_{i,t}^{DG}}{3} &= \sum_{i\in\mathcal{N}_n} x_{n,t}\left( P_{i,t,\phi}^D - P_{i,t,\phi}^{PV} \right) \label{eqn:Pbalance} \\
  \sum_{i\in\mathcal{N}^{ES}} Q_{i,t,\phi}^{ES} + \sum_{i\in\mathcal{N}^{DG}} \frac{Q_{i,t}^{DG}}{3} &= \sum_{i\in\mathcal{N}_n} x_{n,t} Q_{i,t,\phi}^D \label{eqn:Qbalance}
  \end{align}
\end{subequations}

Equations (\ref{eqn:ESPlim})-(\ref{eqn:ESSlim}) represent the real, reactive, and inverter limits of the energy storage devices $\forall i\in\mathcal{N}^{ES}$. In equation-(\ref{eqn:ESSlim}), $\gamma_t \le 1$ indicates the reserve factor imposed on the inverter which reduces the actual limits of the inverter to account for forecast error and other modeling errors. Further, (\ref{eqn:ESSlim}) is a quadratic constraint which is linearized using the m-sided polygon technique with m = 6 as proposed in \cite{ahmadi2015}.
\begin{subequations}\label{eqn:ESPower}
  \begin{align}
  -x_{n,t}\underline{P}_i^{ES} \le \sum_{\phi\in\Phi}P_{i,t,\phi}^{ES} &\le  x_{n,t}\overline{P}_i^{ES} \label{eqn:ESPlim} \\
  0 \le  \sum_{\phi\in\Phi}Q_{i,t,\phi}^{ES} &\le x_{n,t}\overline{Q}_i^{ES} \label{eqn:ESQlim}\\
  \left(\sum_{\phi\in\Phi}P_{i,t,\phi}^{ES}\right)^2 + \left(\sum_{\phi\in\Phi}Q_{i,t,\phi}^{ES}\right)^2 &\le x_{n,t} \left(\gamma_t\overline{S}^{ES}_i\right)^2 \label{eqn:ESSlim}
  \end{align}
\end{subequations}

Equations (\ref{eqn:SoC})-(\ref{eqn:SoClim}) represent the SoC constraints of the battery $\forall i\in\mathcal{N}^{ES}$. The temporal change in SoC based on the expected output of ES is shown in (\ref{eqn:SoC}) and the absolute SoC limits is defined in (\ref{eqn:SoClim}) where $\eta_i$ is the efficiency of $i^{th}$ ES unit.
\begin{subequations}\label{eqn:ESSoC}
  \begin{align}
    SoC_{i,t} &= SoC_{i, t-1} - \frac{\sum_{\phi\in\Phi}P_{i,t,\phi}^{ES}}{\overline{E}_i^{ES}\eta_i^{ES}}\Delta t \label{eqn:SoC}\\
    \underline{SoC}_i &\le SoC_{i,t} \le \overline{SoC}_i \label{eqn:SoClim}
  \end{align}
\end{subequations}

Equations (\ref{eqn:DGP})-(\ref{eqn:DGFuellim}) represent the diesel generator constraints $\forall i\in\mathcal{N}^{DG}$. Equations (\ref{eqn:DGP}) represents the real power limits of the DG which is dispatched in fixed power factor mode with power factor angle  $\theta_i$ as indicated in (\ref{eqn:DGQ}). The temporal change of fuel based on expected dispatch of diesel generator is shown in (\ref{eqn:DGFuel}) and the fuel limits are imposed using (\ref{eqn:DGFuellim}).
\begin{subequations}\label{eqn:DG}
  \begin{align}
    x_{i,t} \underline{S}_i^{DG} \cos\theta_i &\le P_{i,t}^{DG} \le x_{i,t}\overline{S}_i^{DG}\cos\theta_i \label{eqn:DGP}\\
    Q_{i,t}^{DG} &= P_{i,t}^{DG}\tan\theta_i \label{eqn:DGQ}\\
    F_{i,t} &= F_{i,t-1} - \left(y_{i,t}\alpha_i + \beta_{i}P_{i,t}^{DG}\right)\Delta t \label{eqn:DGFuel}\\
    \underline{F}_i &\le F_{i,t} \le  \overline{F}_i \label{eqn:DGFuellim}
  \end{align}
\end{subequations}

Equation (\ref{eqn:MSD}) represents the minimum service duration (MSD) constraint $\forall n\in\mathcal{N}^{LG}$,  $t\in\mathcal{T}\setminus\{1\}$. When a load group is decided to be picked up by the microgrid, it has to be supplied for a minimum duration of $\alpha^{MSD}$. A constraint to ensure load group connectivity sequence to maintain radiality is also included.
\begin{subequations}\label{eqn:SwitchEquations}
\begin{align}
\sum_{t'=t}^{t+\alpha^{MSD}-1} x_{n,t'} &\ge \alpha^{MSD}(x_{n,t} - x_{n,t-1}) \label{eqn:MSD}
\end{align}
\end{subequations}

Equation (\ref{eqn:MinUpTime}) represents the minimum up time ($\alpha^{up}$) constraint and (\ref{eqn:StartUp-1}), (\ref{eqn:StartUp-2}) together compute the startup cost of diesel generators $\forall i\in\mathcal{N}^{DG}$. Every time the diesel generator is switched on, a start up cost of $C^{up}$ is incurred.
\begin{subequations}\label{eqn:DGStartup}
  \begin{align}
    \sum_{t'=t}^{t+\alpha^{up}-1} y_{i,t'} &\le \alpha^{up}(y_{i,t} - y_{i,t-1}) \label{eqn:MinUpTime}\\
    C_{i,t} &\ge C^{up}\left( y_{i,t} - y_{i,t-1} \right ) \label{eqn:StartUp-1}\\
    C_{i,t} &\ge 0 \label{eqn:StartUp-2}
  \end{align}
\end{subequations}

\subsubsection{Stage-2: Dispatching Problem}\label{subsec:stage2}
This stage utilizes a short-term forecast which is generally more accurate than stage-1 forecast for actual dispatching of the resources and switch control closer to real-time operation. The objective function is shown in (\ref{eqn:stage2obj}) where the first term maximizes the amount of load served with higher priority to critical loads. Since the switch control is re-evaluated, the second term minimizes the deviation of load group status from stage-1 results $ \hat{x}_{n,t}$. The third term minimizes the deviation of diesel generator dispatch from the stage-1 scheduled value $\hat{P}^{DG}_{i,t}$.  Here, $k\in\mathcal{K}_t$ is the dispatch interval  as indicated in figure-\ref{fig:EMSframework}.
\begin{multline}\label{eqn:stage2obj}
\max_{x} \quad \sum_{n\in\mathcal{N}^{LG}}\left[ w_nx_{n,k}\sum_{\phi\in\Phi}P_{i,k,\phi}^D - w_n^{sw}\left[ \hat{x}_{n,t} - x_{n,k} \right]^2\right]\\
- \sum_{i\in\mathcal{N}^{DG}} \left[\hat{P}_{i,t}^{DG} - P_{i,k}^{DG} \right]^2
\end{multline}

The real power balance constraint is similar to stage-1 equation (\ref{eqn:Pbalance}) but defined over $k\in\mathcal{K}_t$. The additional term involving  $\lambda_n$ and  $\hat{\epsilon}_t$ is the forecast correction term which is explained in section-\ref{sec:forecastcorrection}.
\begin{multline}\label{eqn:stage2Pbalance}
  \sum_{i\in\mathcal{N}^{ES}} P_{i,k,\phi}^{ES} + \sum_{i\in\mathcal{N}^{DG}} \frac{P_{i,k}^{DG}}{3} \\= \sum_{n\in\mathcal{N}^{LG}}\left[ \sum_{i\in\mathcal{N}_n} x_{n,k}\left( P_{i,k,\phi}^D - P_{i,k,\phi}^{PV} \right) - \lambda_n \frac{\hat{\epsilon}_t}{3}\right]
\end{multline}

The reactive power constraint in (\ref{eqn:stage2Qbalance}) is similar to real power constraint. Note that there is no forecast error correction term here, as  the major issue in forecast comes from BTM PV (as we will see in section-\ref{sec:forecastcorrection}) and there is no reactive power contribution from PV..
\begin{equation}\label{eqn:stage2Qbalance}
  \sum_{i\in\mathcal{N}^{ES}} Q_{i,k,\phi}^{ES} + \sum_{i\in\mathcal{N}^{DG}} \frac{Q_{i,k}^{DG}}{3} = \sum_{n\in\mathcal{N}^{LG}} \sum_{i\in\mathcal{N}_n} x_{n,k} Q_{i,k,\phi}^D 
\end{equation}

The constraints (\ref{eqn:ESPower}), (\ref{eqn:ESSoC}), and (\ref{eqn:DG}) can be imported directly from stage-1 problem by replacing subscripts $t, \Delta t, \mathcal{T}$ by  $k, \Delta k, \mathcal{K}_t$ to represent the MES and DG constraints.

\section{Enhancements for Robustness}\label{sec:MicrogridSecurity}
The performance of the main proposed scheme depends heavily on the accuracy of the forecast on the net load, and as indicated earlier this forecast can have significant error. The other challenge is the effective scheduling of the limited fuel of the DG to ensure service to loads over multiple days. To address these challenges, and thus to enhance the robustness of the proposed EMS we propose the following strategies:
\begin{itemize}
  \item Multi-Day Fuel Rationing
  \item Learning-based Forecast Correction
  \item Dynamic Reserve Management
\end{itemize}

\subsection{Multi-Day Fuel Rationing}
For the generation mix of this feeder microgrid we consider a MES, a DG, and many distributed BTM PVs. MES serves as the grid forming (GFM) resource due to its  fast dynamic response, and DG serves as the main energy source. As stated earlier in sec-\ref{subsec:stage1}, we solve a receding horizon problem on final day of restoration and rolling horizon problem on all other days with a horizon of 24 hours primarily because smart meter data is available real-time at 15-min interval. Hence, we can forecast the load in a rolling fashion updated every 30-min to 1hr but PV forecast requires weather information which is impossible to obtain in a rolling fashion by the utility. Also it is very challenging to accurately forecast PV irradiance beyond 24 hour period. For this reason we can only solve a rolling horizon problem with fixed horizon of 24 hours initially. Since the resources need to be rationed over multiple-days we propose a multi-day fuel rationing framework that ensures availability of DG over all days of a restoration process even under significant uncertainty in load and PV. We develop a fuel management algorithm that can efficiently ration the fuel over multiple days.
\begin{equation}\label{eqn:FuelManagement}
F_{i,r} = \left.
\begin{cases}
  F_{i,t-1} - \frac{(F_{i,t-1} - F_{f})(\mid\mathcal{T}\mid d + t - 1)}{ \mid \mathcal{T} \mid \cdot  \mid \mathcal{D}  \mid }, & \text {for } d <  \mid \mathcal{D} \mid\\
  F_f & \text{for } d =  \mid \mathcal{D} \mid\\
\end{cases}
\right.
\end{equation}
\begin{equation}\label{eqn:finalreserveconstraint}
  F_{i, \mid \mathcal{T} \mid } \le F_{i,r}
\end{equation}

Equation-(\ref{eqn:FuelManagement}) highlights a piece-wise linear function defined $\forall i\in\mathcal{N}^{DG}$, $t\in\mathcal{T}$. $F_{i,r}$ is the minimum fuel reserve target at the end of each control cycle in stage-1 scheduling problem, which is enforced as a constraint in (\ref{eqn:finalreserveconstraint}). $F_{i,t-1}$ is the initial fuel at start of each control cycle in stage-1 problem,  $F_f$ is the final reserve desired at the end of multi-day restoration, which can be set to a non-zero value to ration some emergency fuel at the end of restoration to supply only critical loads or to resynchronize the distribution system with main feeder by picking up all the loads in the system. This framework can be extended to energy storage devices acting as grid following (GFL) resources by implementing the same reserve function on SoC of the batteries.

\subsection{Learning-based Forecast Correction}\label{sec:forecastcorrection}
The major issue with BTM PV is lack of real-time data. When individual houses are net metered, it is easier to forecast
the net load but dis-aggregating PV from load and forecasting just the PV component becomes challenging. For the EMS scheme, accurate forecast of the net load variation for each load group is needed. The actual load patterns is usually quite predictable, it is the PV variability that is challenging to forecast for each load group.  We have adopted an effective PV forecast method proposed in \cite{li2021b} for this purpose. The method  uses the irradiance forecast and the PV rating at each house to estimate the BTM PV output. The low accuracy on individual house forecast can become large when aggregated to feeder-level \cite{erdener2022}. Handling such large errors becomes very challenging since we have a MES with limited capacity.
\begin{figure}[htpb]
  \centering
  \includegraphics[width=3.5in]{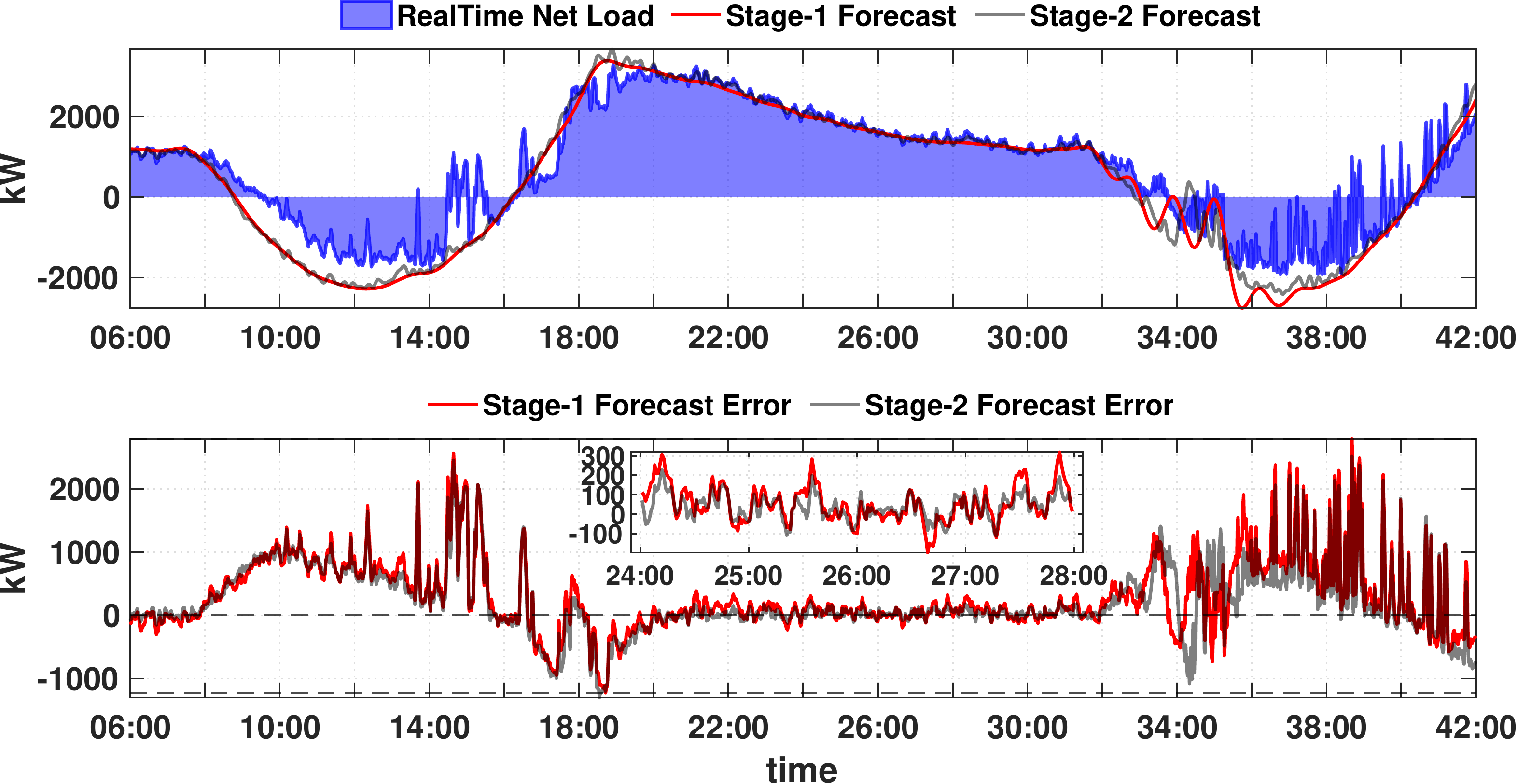}
  \caption{High forecast error in short-term and day-head forecast of aggregated BTM PV. \textit{(top)} Real-time net load measurement, stage-1, and stage-2 forecast. \textit{(bottom)} Total net load forecast error in stage-1 and stage-2 forecast. Zoomed plot shows stage-2 forecast more accurate than stage-1 forecast during no PV duration and error is close to zero.}
  \label{fig:forecastError}\vspace{-0.6cm}
\end{figure}

Figure \ref{fig:forecastError} shows the error in aggregated netload forecast which increases significantly during the presence of PV (6:00 - 18:00 on day-1 and day-2). There are two characteristic to the forecast error introduced by the PV. One is the average forecast error due to error in PV rating assumed, unaccounted rooftop PV systems and modeling errors. The other characteristic is the instantaneous random errors introduced by cloud cover events. The errors can be as high as 2 MW on a 3.5 MW circuit. 

To address the average forecast error issue we introduce the learning-based forecast correction scheme. The scheme uses the battery SOC to estimate the forecast error in PV. This is based on the following relationship between the two quantities:

\begin{equation}
  \Delta P^{D}_t \approx \kappa\Delta SoC_{j, t+1}
\end{equation}
\begin{thm}\label{thm:SoC}
  In a islanded microgrid with battery as a slack generator and  when the proposed rolling optimization framework is used for energy management, then the difference between the SOC computed at stage-1 and real-time SOC measurement at each dispatch point given by $\Delta SoC_{j,t+1}$ is approximately equal to the sum of average forecast error, average power loss and average modeling error given by $\Delta P^{D}_t$ in the dispatch interval $\Delta t$ with a constant $\kappa = \frac{\overline{E}^{ES}\eta}{\Delta t}$.
\end{thm}

This relationship is derived in the appendix. Hence, using this relationship, we can use this estimated forecast error component from previous iteration to adjust the stage-2 forecast error in the next control cycle as shown in (\ref{eqn:Adjust}). Here, $K$ is the historical window length over which the average is taken.
\begin{equation}\label{eqn:Adjust}
\hat{\epsilon}_t = \frac{\sum_{i=t-K}^{t-1} \Delta P^D_i}{K}
\end{equation}

This average value $ \hat{\epsilon}_t$ is then normalized to each load group based on their connected load (given by $\lambda_n$) and subtracted from the stage-2 forecast as shown in (\ref{eqn:stage2Pbalance}).

\subsection{Dynamic Reserve Management}\label{sec:reserve}
Apart from the large average forecast error during PV in fig-\ref{fig:forecastError}, the other issue is the large fast variations caused by cloud cover events which is difficult to forecast even with accurate weather information. Since these are random events, we cannot modify the forecast values to account for these errors as we did in sec \ref{sec:forecastcorrection} but instead we can prepare the system to be robust against such events.

A common method to account for unknown errors in the system while dispatching is to use a power reserve on the generation resource, which is the MES in this case. The reserve provides a head room to supply unknown forecast errors and modeling errors that are not accounted for in the dispatch. This is incorporated as $\gamma_t$ in (\ref{eqn:ESSlim}) where, $\gamma_t\in[0, 1]$ is the reserve factor for the MES which is responsible for power balance in the microgrid.  Even though reserves are efficient in operating the microgrid securely, a large reserve can significantly affect the number of customers served. 

To handle the forecast errors introduced by cloud cover events optimally we propose a dynamic reserve management strategy. The first step is to estimate the random errors from available data. We use the previous historical forecast errors estimates $\Delta P^D_{[t-q, t]}$ to estimate the random spikes using a moving average (MA) model as shown in (\ref{eqn:MAmodel}). The parameters of MA model can be estimated from forecast error estimates obtained from previous few hours.
\begin{equation}\label{eqn:MAmodel}
  \hat{\Delta P^{D}_{t+1}} = \mu + \sum_{i=1}^{q} \theta_i\Delta P_{t-i}^D + \Delta P_{t}^D
\end{equation}

The algorithm to estimate the required reserve given in alg-\ref{alg:DRM} is based on comparing previous interval net load forecast $\hat{P}^{net}_t$ given in (\ref{eqn:netload}) to the actual power measurement at PCC given by $P^{net}_t$ which is the actual total net load in the circuit.
 \begin{equation}\label{eqn:netload}
  \hat{P}^{net}_t = \sum_{i\in\mathcal{N}}^{} \left(P^{D}_{i,t} - P^{PV}_{i,t}\right)
\end{equation}

High reserve requirements are required during two conditions, when actual load is more than expected and when actual PV is lower than expected. In both these conditions net load of microgrid is more than expected. When net load is lower than expected a high reserve will hinder the optimality of the solution. This is further validated in our case study in sec \ref{subsec:perfeval}. So it is critical to dynamically modify the reserve to securely and optimally restore the loads. By comparing the forecasted net load and the actual measured total net load from previous control cycle, we can decide the level of reserve requirement as shown in algorithm \ref{alg:DRM}. This is based on the assumption that net load behavior in the next control cycle $\Delta t$ will be approximately same as previous control cycle $\Delta t$ following a trend. When the trend is higher net load than expected then the reserve is dynamically modified depending on expected forecast error and left at minimum value $\alpha$ in other scenarios.

Even though this method will address the dynamic forecast errors to some extent, we need a back up protection like under frequency load trip or other load shedding algorithm to securely operate the microgrid since it is very difficult to predict the instantaneous cloud cover events from average power measurements.
 \begin{algorithm}[htbp]
 \caption{Dynamic Reserve Management}
 \label{alg:DRM}
 \begin{algorithmic}[1]
 \renewcommand{\algorithmicrequire}{\textbf{Input:}}
 \renewcommand{\algorithmicensure}{\textbf{Output:}}
 \REQUIRE Net load forecast $\hat{P}^{net}_{t}$, Net load measurements  $P^{net}_{t}$, and Forecast error estimate $\Delta \hat{P}^D_{t+1}$
 \ENSURE  Power Reserve Factor $\gamma_t$
  \FOR {$t = k$ to $ \mid \mathcal{H} \mid $}
  \IF {($P^{net}_t > \hat{P}^{net}_t$)}
  \STATE $\gamma_t = (1-\Delta \hat{P}^D_{t+1})$
  \ELSE
  \STATE $\gamma_t = 1-\alpha$, where $\alpha$ is minimum reserve desired.
  \ENDIF
  \STATE $\gamma_t = \mathbb{P}[\gamma_t]$ where $\mathbb{P}$ operator thresholds any input to the set $[\underline{\gamma_t}, \overline{\gamma_t}]$
  \RETURN $\gamma_t$
  \ENDFOR
 \end{algorithmic}
 \end{algorithm}

\section{Case Study and Performance Evaluation}\label{sec:casestudy}
\begin{figure}[htpb]
  \centering
  \includegraphics[width=3.5in]{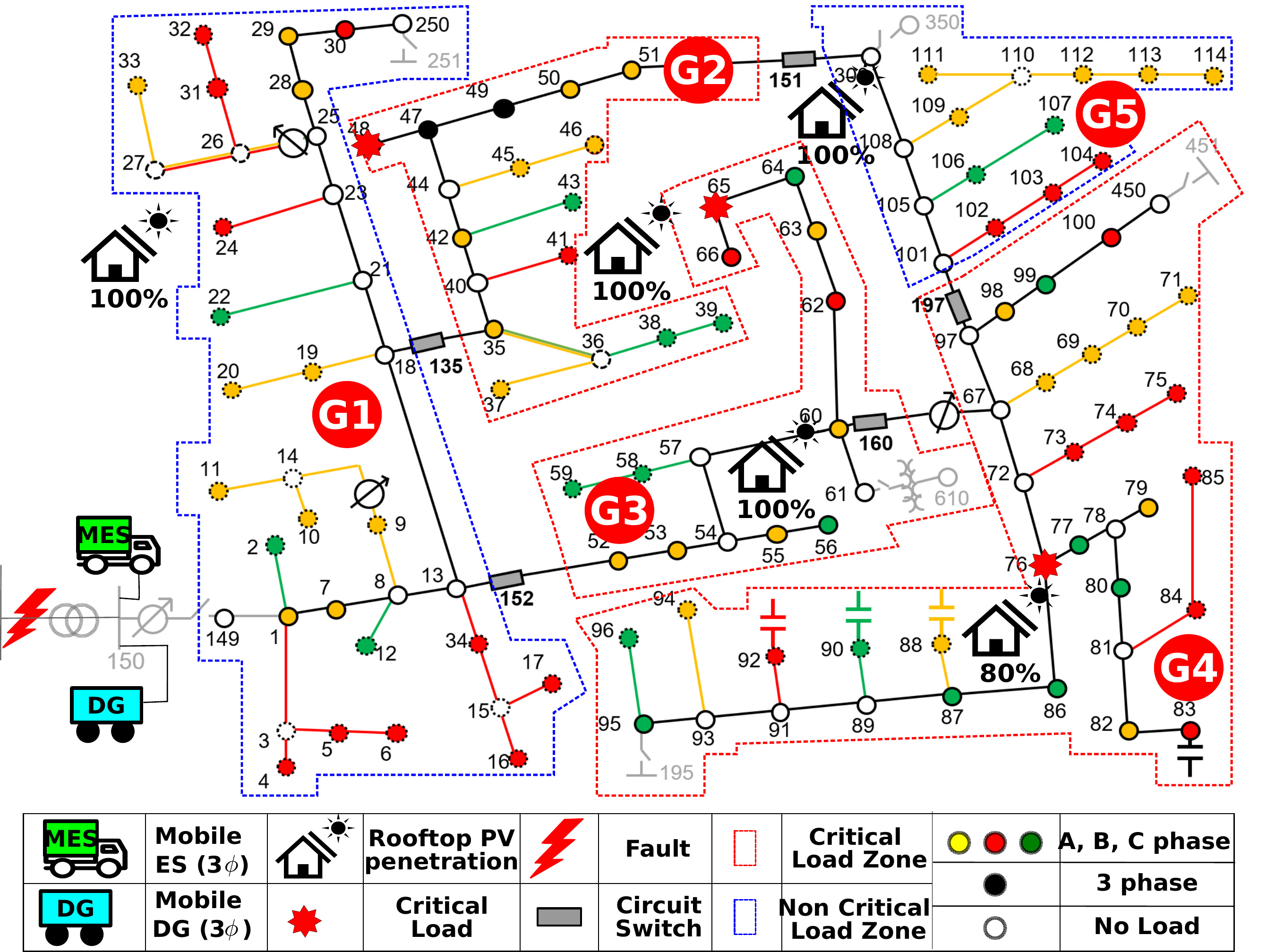}
  \caption{Modified IEEE 123 Node Test Feeder}
  \label{fig:IEEE123}\vspace{-0.6cm}
\end{figure}

To illustrate the performance of the proposed energy management scheme a case study is conducted. The sample system is the IEEE 123 node system shown in fig-\ref{fig:IEEE123}. The power to the system is assumed to be unavailable due to outage and to supply power to the loads during outage, utility brings in a mobile MES and a DG to form the microgrid. Distribution feeder has 5 controllable switches and they divide the feeder into 5 load groups as indicated on the figure.  Really high PV penetration of 80-100\% is considered in each load zone as shown in fig-\ref{fig:IEEE123}. The system resources are summarized in table-\ref{tab:Resources} and the simulation parameters are given in table-\ref{tab:simParam}.
\begin{table}[htpb]
  \centering
  \caption{Location of Resources}\label{tab:Resources}
  \begin{tabular}{c|c|c|c}
    \hline
    Resource & Location & Rating (kW) & \begin{tabular}{@{}c@{}}Rating\\ (L/kWh)\end{tabular}\\
    \hline
    \hline
    \begin{tabular}{@{}c@{}}Critical\\ Load Nodes\end{tabular} & 48, 65, 76 & 210, 140, 245 kW& - \\
    \hline
    Battery & Substation & 2000 kW& 8000 kWh\\
    \hline
    \begin{tabular}{@{}c@{}}Diesel\\Generator\end{tabular} & Substation & 4000 kW & 10000 L\\
    \hline
    Rooftop PV & All nodes & 19 - 111 kW & - \\
    \hline
  \end{tabular}
\end{table}

\begin{table}[htpb]
  \centering
  \caption{Simulation Parameters}
  \label{tab:simParam}
  \begin{tabular}{c|c|c|c}
  \hline
   Parameter & Value & Parameter & Value \\
   \hline
   \hline
   $\Delta t$, $\Delta k$, $\Delta h$ & 30, 5, 1 min  & $\mathcal{T}$, $\mathcal{K}_t$ & 24 hr, 30 min\\
   \hline
   $w_1$,$w_5$ & 0.01 & $\alpha^{up}$ & 1 hr\\
   \hline
   $w_2$,$w_3$,$w_4$ & 0.4, 0.3, 0.2 & $\alpha^{MSD}$ & 2 hr\\
   \hline
   $\alpha_i$, $\beta_i$ $\forall i \in \mathcal{N}^{DG}$ & 84.87, 0.20 &  $C^{up}$ & 6\\
   \hline
  \end{tabular}
\end{table}

\begin{figure}[htpb]
  \centering
  \includegraphics[width=3.5in]{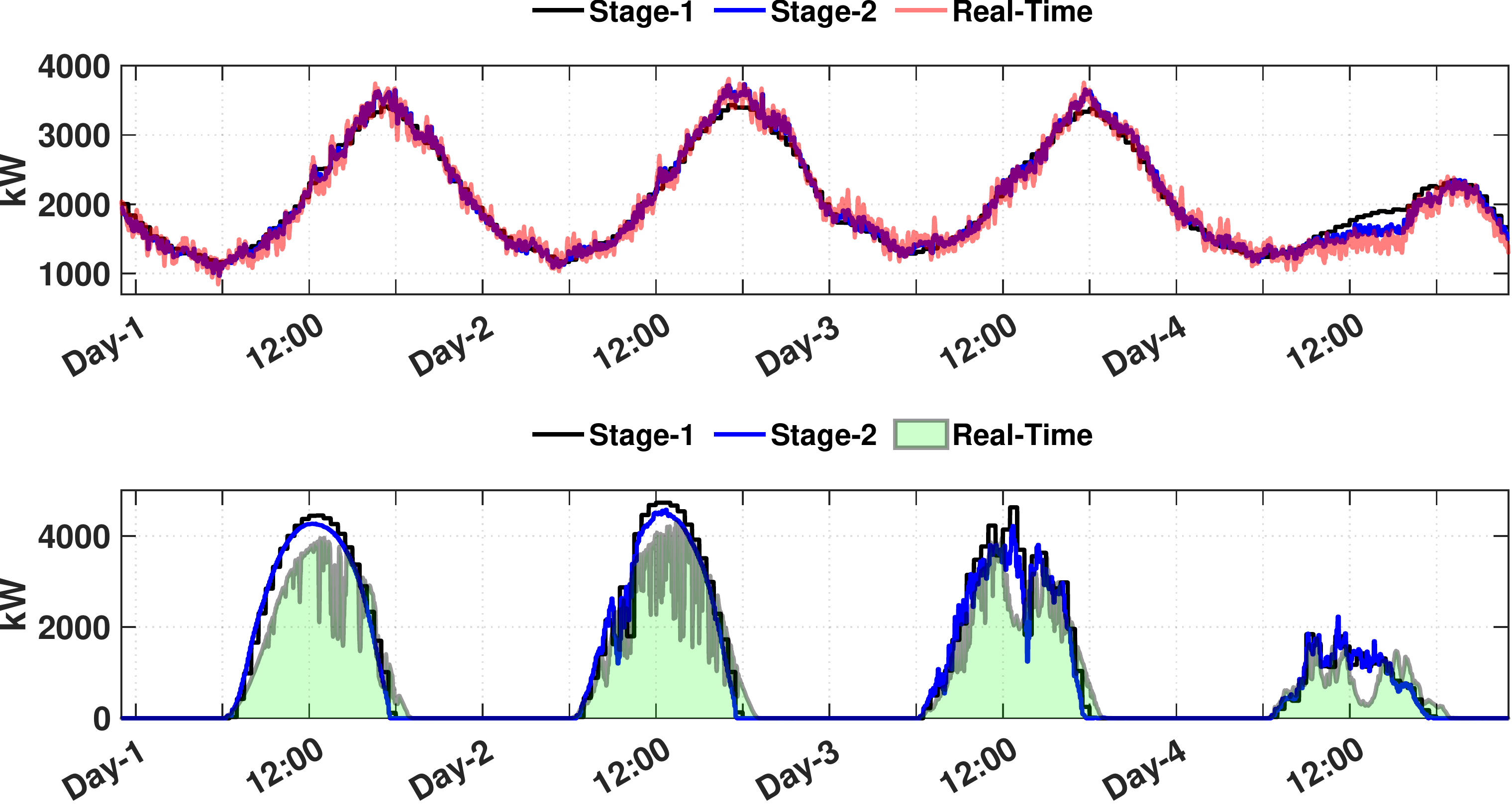}
  \caption{Total Load Forecast (\textit{top}), and PV Forecast (\textit{bottom}) against real-time data for 4 days.}
  \label{fig:forecastErr}
\end{figure}

The outage is assumed to be for a 48 hour period starting at midnight. The forecast of load and PV for 4 day period is shown in fig. \ref{fig:forecastErr}. The forecast are obtained using the algorithm proposed in \cite{li2021b} trained on pecan street summer data for load and corresponding weather data from nearby weather station in Texas for irradiance forecast. As indicated in the figure, stage-1 load forecast are obtained in a rolling manner every 30-min using smart meter data while stage-1 irradiance forecast is obtained in a day-ahead stage since there is no real-time data to update the forecasts. The day ahead forecast is repeated to obtain a rolling forecast. Stage-2 load and irradiance are short term forecast obtained closer to real-time.
The total installed capacity of PV at each node is utilized to get the final stage-1 and stage-2 PV forecast from irradiance information. The following inference can be made from the load and PV forecast results in fig. \ref{fig:forecastErr}:
\begin{itemize}
  \item The forecast error in load is small due to availability of real-time smart meter data.
  \item Stage-2 forecast of load is much closer to real-time data compared to stage-1 forecast. While, short-term stage-2 forecast in PV provides no improvement from stage-1 forecast due to unavailability of real-time information to update the forecast.
  \item The PV is consistently over estimated since rating of PV might be smaller than actual installed capacity which leads to significant error when multiplied with irradiance forecast. Also, one irradiance profile is used to obtain the output of all PV units which leads to further error in forecast.
  \item The cloud cover events are difficult to forecast from weather data and eventually shows up as instantaneous random forecast errors in net load.
\end{itemize}

The simulations are carried out using a PC with Intel Core i7-11700F CPU @ 4.8 GHz processor and 64 GB RAM. The prposed schemes are implemented and solved in Matlab using the Yalmip environment with GUROBI 9.5 solver. The microgrid is simulated using OpenDSS and it is interfaced  with Matlab using the COM Interface. The simulations are run for 48 hours starting at Day-1 in fig. \ref{fig:forecastErr}.

To assess the performance of the proposed scheme, the following performance metrics are used: $P^{CL}_{\%}$ and  $P^{NCL}_{\%}$ are the percentage of critical and non-critical loads (NCLs) served. $P^{PV}_{\%}$ is the percentage of available PV utilized during restoration. $T^{CL}$ and $T^{NCL}$ are the average served duration of CL and NCL. $N^{CL}$ is the number of interruptions in serving critical loads (CLs) and $N^{\mu G}$ is the number of times microgrid is shutdown. $I^{\mu G}$ is the total duration microgrid was shutdown due to unavailability of resourcses.

\subsection{Performance Evaluation}\label{subsec:perfeval}
To demonstrate the effectiveness of the forecast error adjustment and dynamic reserve management strategies, we considered 4 different cases:
\begin{itemize}
  \item \textbf{Base Case}: Fixed power reserve of 400kW ($\gamma_t = 0.8$), no forecast error correction is included to correct the stage-2 forecast, and fixed target fuel reserve through out two day restoration $F_{i,r} = 500L$.
  \item \textbf{Case-1}: Fixed power reserve of 400kW ($\gamma_t = 0.8$) through out the two day restoration period. But, stage-2 forecast is corrected in every stage-1 control cycle based on historical estimation using a window length of 5 hrs. Fuel reserve management as per eqn-(\ref{eqn:FuelManagement}).
  \item \textbf{Case-2}: 20\% of current stage-1 net load forecast is kept as reserve in stage-2, no forecast error correction included to correct the stage-2 forecast, and fuel reserve management as per eqn-(\ref{eqn:FuelManagement}).
  \item \textbf{Case-3}: Reserve is adjusted dynamically based on algorithm-\ref{alg:DRM}. Stage-2 forecast is corrected similar to case-2. Fuel reserve management as per eqn-(\ref{eqn:FuelManagement}).
\end{itemize}

 The 48 hour restoration is simulated using the proposed management scheme and  using the load and PV profiles shown in fig-\ref{fig:forecastErr}. The results are summarized in table-\ref{tab:caseStudy}. To evaluate the performance we introduce additional metrics such as $N^{\mu G}_{UnSch}$ which is the number of unique unscheduled shutdown of microgrid due to under frequency load shedding which is assumed to last for 30 min and the total duration of such unscheduled shutdown is given by $T^{\mu G}_{UnSch}$. $T^{\mu G}_{Total}$ is the sum of total scheduled $T^{\mu G}_{Sch}$ and unscheduled $T^{\mu G}_{UnSch}$ shutdown of the microgrid.
 \vspace{-0.4cm}
\begin{table}[htbp]
\caption{Different cases to evaluate forecast error correction and dynamic reserve management module}
\label{tab:caseStudySetup}
\begin{tabular}{c|c|l|c}
\hline
\textbf{Cases} & \textbf{\begin{tabular}[c]{@{}c@{}}Error\\ Correction\end{tabular}} & \multicolumn{1}{c|}{\textbf{Fuel Management}} & \textbf{\begin{tabular}[c]{@{}c@{}}Reserve\\ $1-\gamma_t$\end{tabular}} \\
\hline\hline
Base Case         & No                                                                           & Fixed $F_{i,r} = 500L$                        & 400 kW                                                                  \\
Case-1         & Yes                                                                          & As per eqn (\ref{eqn:FuelManagement})                        & 400 kW                                                                  \\
Case-2         & No                                                                           & As per eqn (\ref{eqn:FuelManagement})                & $0.2\sum(\hat{\boldsymbol{P}}^D - \hat{\boldsymbol{P}}^{PV})$                           \\
Case-3         & Yes                                                                          & As per eqn (\ref{eqn:FuelManagement})                                   &  as per algorithm-\ref{alg:DRM}                                                                      \\\hline
\end{tabular}
\end{table}
\vspace{-0.4cm}

\subsection{Fuel Management}\label{subsec:fuelmanagemnet}
\begin{figure}[htpb]
  \centering
  \includegraphics[width=3.5in]{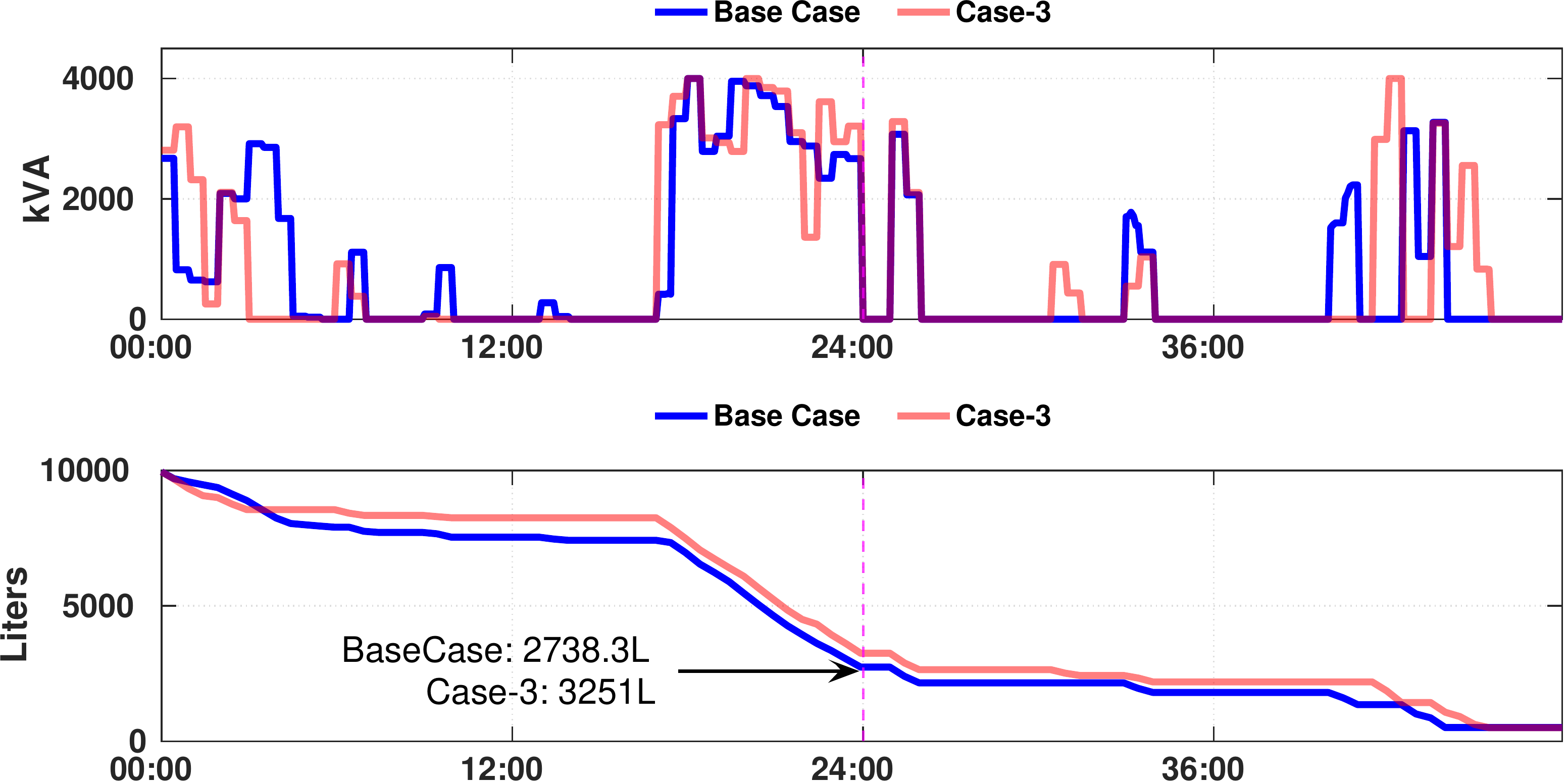}
  \caption{(\textit{top}) Comparison of DG dispatch between different fuel management schemes. (\textit{bottom}) Comparison of DG Fuel usage over two day restoration with different fuel management schemes.}
  \label{fig:fuelUsage}\vspace{-0.5cm}
\end{figure}

Figure-\ref{fig:fuelUsage} shows the actual usage of fuel over a two day period for day-1 simulation between base case with fixed reserve and case-4 with proposed fuel management and corresponding DG output. The DG is predominantly used to either charge the battery while serving low load or to completely serve the peak load in the circuit. From the figure it is evident that the proposed fuel scheme ensures higher fuel reserve for second day of restoration compared to fixed fuel reserve scheme. This directly leads to about 7\% increase in critical load served which is highlighted in table-\ref{tab:caseStudy}. The duration of critical loads served is also considerably increased. This shows the importance of mulit-day fuel rationing when only a day-ahead forecast is available.

From the table we see that critical loads are not served 100\%, this doesn't mean the loads are under outage during the restoration process. Since this EMS is from a utility perspective, utility will communicate the hours critical loads will not be served as scheduled outages from scheduling framework and the critical loads will use the local generation during these outage hours. Nevertheless, we want to minimize this duration since the local generation at critical loads can be limited or unavailable.

\subsection{Forecast Error Adjustment and Reserve Management}\label{subsec:reservemanagement}
In table-\ref{tab:caseStudy}, both on day-1 and day-3 we see that Case-2 with forecast error adjustment performs extremely well across all metrics compared to Case-1 without any error adjustment. There is both significant increase in load served and PV utilized during restoration which highlights the importance of forecast error adjustment. Both the cases suffer from unscheduled outages due to violation of battery limits caused by high forecast error which leads to about 2 to 3 hours of unscheduled shutdown on day-1.  Unscheduled outages are a inconvenience to the customer, especially to critical loads, and hence it is desirable to reduce such unscheduled outages during restoration .
\begin{figure}[htpb]
  \centering
  \includegraphics[width=3.5in]{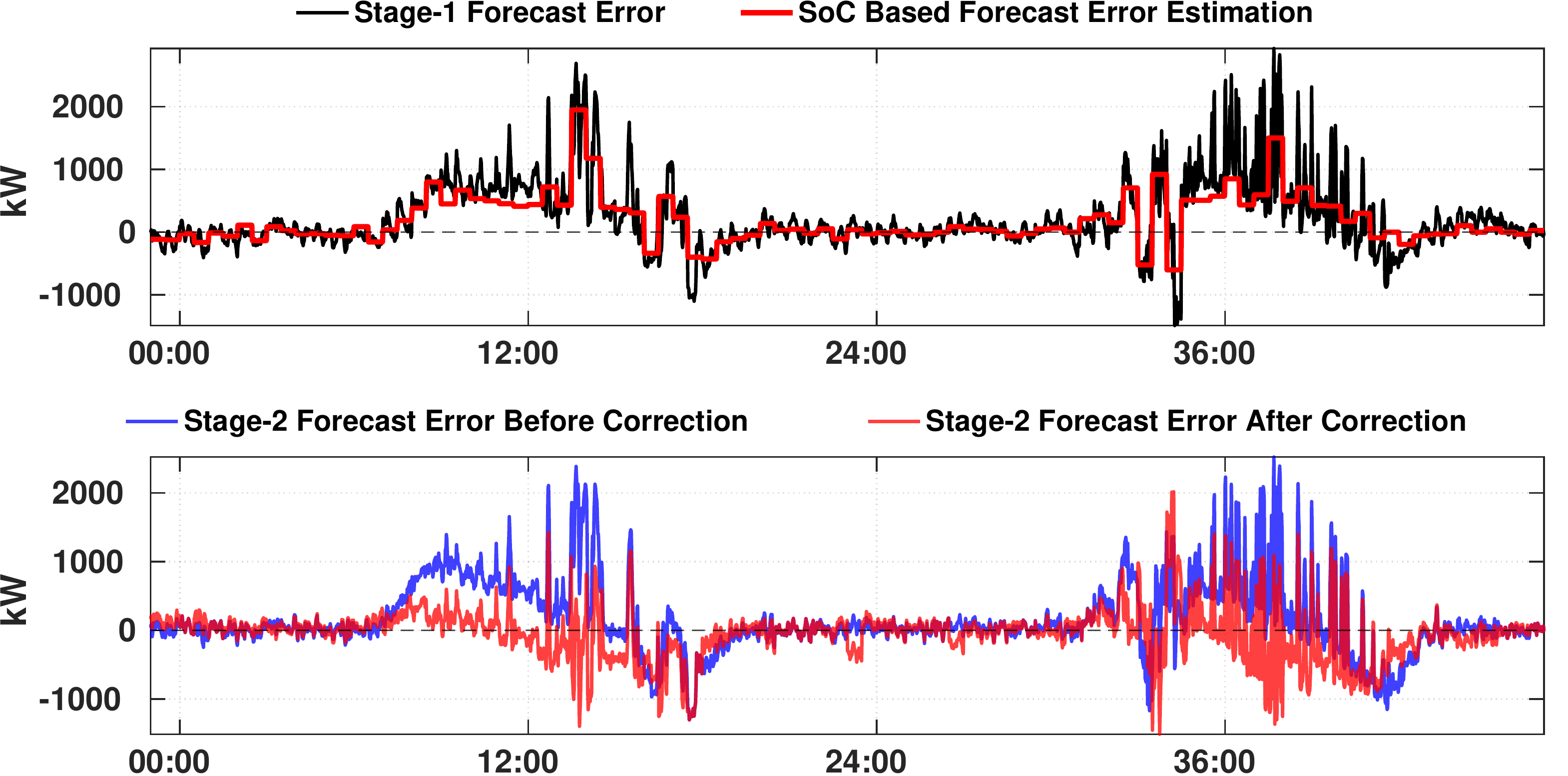}
  \caption{(\textit{top}) Average Net Load Forecast Error Estimation from Battery SoC on Day-1 48 hr restoration. (\textit{bottom}) Comparison of Stage-2 net load forecast error before and after correction using SoC based average forecast error estimation.}
  \label{fig:day1Est}
\end{figure}
\begin{table*}[htpb]
\centering
\caption{Case study of Energy Management with different robustness enhancement modules over multiple days}
\label{tab:caseStudy}
\begin{tabular}{l|cccc||cccc}
\hline
\multicolumn{1}{c|}{\multirow{2}{*}{Metric}} & \multicolumn{4}{c||}{Day-1}                                                                           & \multicolumn{4}{c}{Day-3}                                                                           \\ \cline{2-9} 
\multicolumn{1}{c|}{}                        & \multicolumn{1}{c|}{Base Case}  & \multicolumn{1}{c|}{Case-1}  & \multicolumn{1}{c|}{Case-2}  & Case-3  & \multicolumn{1}{c|}{Base Case}  & \multicolumn{1}{c|}{Case-1}  & \multicolumn{1}{c|}{Case-2}  & Case-3  \\ \hline\hline
$P^{CL}_{\%}$                                 & \multicolumn{1}{c|}{72.73\%} & \multicolumn{1}{c|}{78.13\%} & \multicolumn{1}{c|}{76.78\%} & \textbf{79.2}\% & \multicolumn{1}{c|}{71.69\%} & \multicolumn{1}{c|}{75.59\%} & \multicolumn{1}{c|}{72.59\%} & \textbf{76.44}\% \\ 
$P^{NCL}_{\%}$                                & \multicolumn{1}{c|}{71.1\%} & \multicolumn{1}{c|}{74.63\%} & \multicolumn{1}{c|}{72.88\%} & \textbf{75.244}\%  & \multicolumn{1}{c|}{67.76\%} & \multicolumn{1}{c|}{70.55\%} & \multicolumn{1}{c|}{68.45\%} & \textbf{71.38}\% \\
$P^{PV}_{\%}$                                 & \multicolumn{1}{c|}{78.33\%}  & \multicolumn{1}{c|}{86.06\%} & \multicolumn{1}{c|}{79.63\%} & \textbf{86.92}\% & \multicolumn{1}{c|}{81.88\%} & \multicolumn{1}{c|}{88.27\%} & \multicolumn{1}{c|}{83.68\%} & \textbf{90.35}\% \\
$T^{\mu G}_{sch}$                             & \multicolumn{1}{c|}{2h 40m}  & \multicolumn{1}{c|}{\textbf{1h 30m}}  & \multicolumn{1}{c|}{2h 10m}  & 2h 30m  & \multicolumn{1}{c|}{4h 5m}   & \multicolumn{1}{c|}{4h 5m}   & \multicolumn{1}{c|}{3h 30m}  & \textbf{3h 30m}  \\
$N^{\mu G}_{sch}$                             & \multicolumn{1}{c|}{4}       & \multicolumn{1}{c|}{4}       & \multicolumn{1}{c|}{4}       & \textbf{3}       & \multicolumn{1}{c|}{6}       & \multicolumn{1}{c|}{5}       & \multicolumn{1}{c|}{5}       & \textbf{2}       \\
$T^{CL}$                                      & \multicolumn{1}{c|}{37h 15m} & \multicolumn{1}{c|}{38h 40m} & \multicolumn{1}{c|}{38h 15m} & \textbf{39h 20m} & \multicolumn{1}{c|}{34h 45m} & \multicolumn{1}{c|}{36h 5m}  & \multicolumn{1}{c|}{35h 25m} & \textbf{37h 5m}  \\
$N^{CL}$                                      & \multicolumn{1}{c|}{8}       & \multicolumn{1}{c|}{\textbf{8}}       & \multicolumn{1}{c|}{8}       & 9       & \multicolumn{1}{c|}{10}      & \multicolumn{1}{c|}{8}       & \multicolumn{1}{c|}{9}       & \textbf{7}       \\
$T^{\mu G}_{UnSch}$                           & \multicolumn{1}{c|}{1h}  & \multicolumn{1}{c|}{30m}     & \multicolumn{1}{c|}{2h}      & \textbf{0}       & \multicolumn{1}{c|}{1h 30m}  & \multicolumn{1}{c|}{1h}      & \multicolumn{1}{c|}{1h 30m}  & \textbf{30m}     \\
$N^{\mu G}_{UnSch}$                           & \multicolumn{1}{c|}{2}       & \multicolumn{1}{c|}{1}       & \multicolumn{1}{c|}{4}       & \textbf{0}       & \multicolumn{1}{c|}{3}       & \multicolumn{1}{c|}{2}       & \multicolumn{1}{c|}{3}       & \textbf{1}       \\
$T^{\mu G}_{Total}$                           & \multicolumn{1}{c|}{3h 40m}  & \multicolumn{1}{c|}{\textbf{2h 10m}}  & \multicolumn{1}{c|}{4h 10m}  & 2h 30m  & \multicolumn{1}{c|}{5h 35m}  & \multicolumn{1}{c|}{5h 5m}   & \multicolumn{1}{c|}{5h}      & \textbf{4h}      \\ \hline
\end{tabular}
\end{table*}

The reason for better performance of Case-2 is the forecast error adjustment strategy. Figure-\ref{fig:day1Est} shows the forecast error estimated from SoC measurements compared against the actual forecasts. This figure highlights that the estimation strategy is good at estimating the average forecast error but cannot account for the instantaneous errors. Also, after adjustment the stage-2 forecast error is close to zero compared to before correction. The high average error during PV is captured in this case and by adjusting the forecast we are able to absorb about 8\% higher PV on day-1, which eventually leads to increase in load served and better performance. The percentage increase on day-3 results is lower because the actual PV available on day-4 being a cloudy day is low. 

Case-3 is a simple reserve management strategy where we keep a percentage of net-load forecast as reserve with an intuition that high reserve is required at high PV and high load duration which are mutually exclusive. But Case-3 seems to perform much poorer than Case-2 which just has fixed reserve through out the restoration. Having a high reserve hinders optimality which is visible in the percentage of load served and PV absorbed but such high reserve in case-3 must reduce the number of unscheduled outages which it fails to do so compared to case-2. For this purpose we use the proposed dynamic reserve management strategy (Case-4) which retains the benefits (day-1) and sometimes performs better (day-2) than Case-2 in terms of percentage load served and PV absorbed but the main benefit of case-4 is the reduction in number and duration of unscheduled outages. It is important to note that proposed strategy cannot completely eliminate the unscheduled outages because in some instances the forecast error can be so high that limited resources cannot handle such errors through reserves and we rely on under-frequency load shedding during these events.
\begin{figure}[htpb]
  \centering
  \includegraphics[width=3.5in]{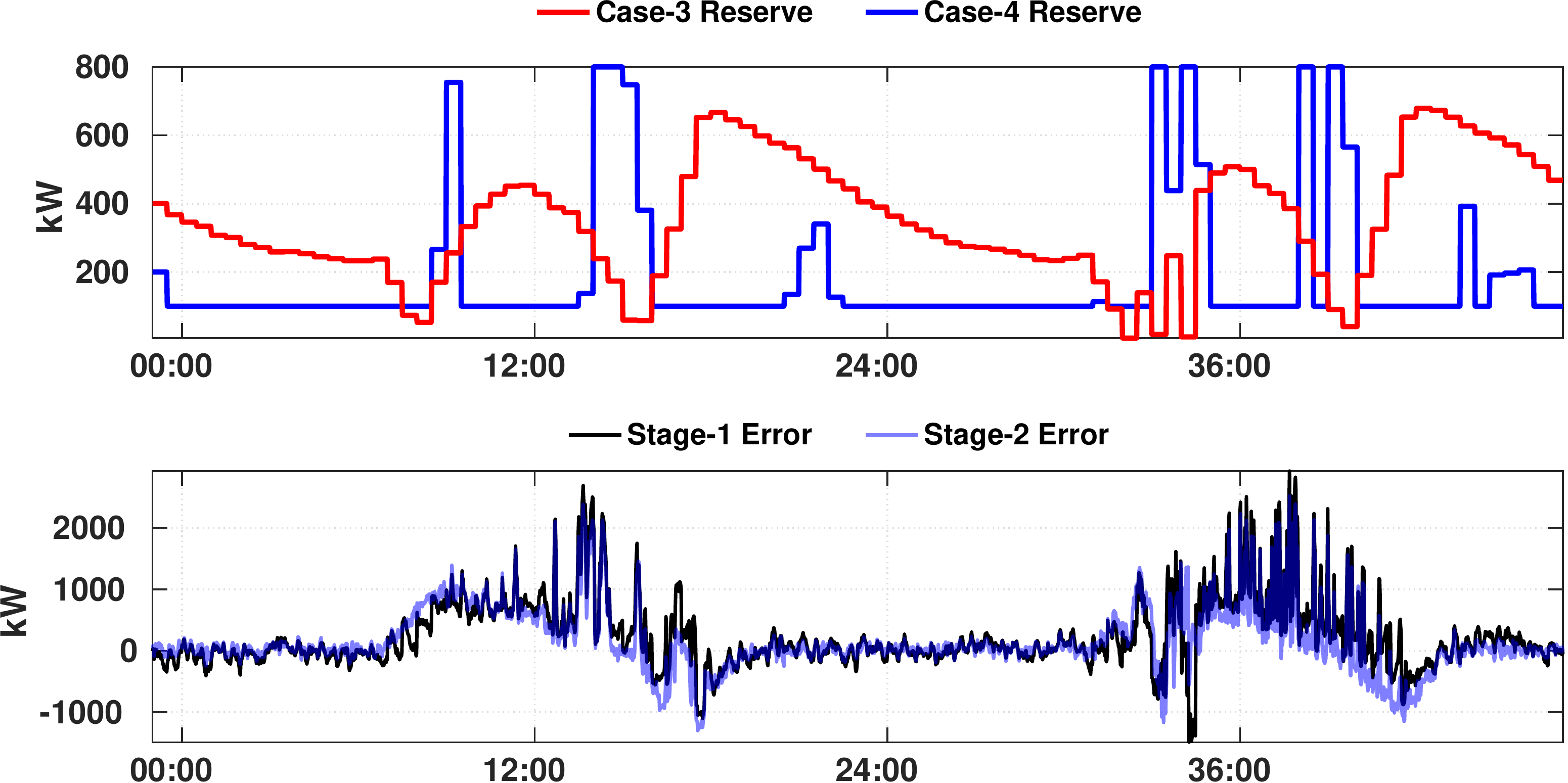}
  \caption{Dynamic reserve adjustment (\textit{top}) compared against actual forecast error (\textit{bottom})}
  \label{fig:reserve}
\end{figure}

The reserves of case-3 and case-4 are shown in fig-\ref{fig:reserve}. Case-3 keeps a high reserve even during low forecast error regions between 18:00-6:00 which hinders the performance of the algorithm. Also, during cloud cover events with high spikes the reserve is not high enough which leads to unscheduled outages. In contrast case-4 keeps the reserve to minimum during low forecast error regions (no PV) and a high reserve only during significant cloud cover events that happens during net load zero region which is the main cause of battery violations since diesel generator is not dispatched during this low net load conditions. Hence, the algorithm intelligently keeps a high reserve around 9:00 and 16:00 where net load is zero. The other cloud cover events around 12:00 noon is not significant since the actual forecast of PV is higher than load and keeping high reserve will only hinder optimality.

\section{Conculsion}\label{sec:conclusion}
This paper proposes a two-stage intelligent energy management strategy for a feeder microgrid to be deployed on a distribution feeder to provide service during a long outage.  To assure that the proposed method has robust performance, three special schemes have been introduced to addresses the unique challenges associated with management of these microgrids:  The multi-day fuel rationing scheme assures that the load in the circuit is served over multiple days of the restoration which increases the overall service to critical loads and improves PV utilization. The novel state-of-charge based forecast error estimation strategy which learns the forecast error online and adjusts the near real-time forecast significantly increases the BTM PV utilization during the restoration. The dynamic reserve management strategy introduced to  handle sudden cloud cover events that can lead to unscheduled outages of the microgrid. A case study on a sample system illustrates the effectiveness of the proposed scheme, as it shows that the proposed error correction based reserve adjustment is quite effective even when there is high variability in PV and considerably reduces the number of unscheduled outages during the restoration process.

\bibliography{references.bib}

\begin{thebibliography}{10}
\providecommand{\url}[1]{#1}
\csname url@samestyle\endcsname
\providecommand{\newblock}{\relax}
\providecommand{\bibinfo}[2]{#2}
\providecommand{\BIBentrySTDinterwordspacing}{\spaceskip=0pt\relax}
\providecommand{\BIBentryALTinterwordstretchfactor}{4}
\providecommand{\BIBentryALTinterwordspacing}{\spaceskip=\fontdimen2\font plus
\BIBentryALTinterwordstretchfactor\fontdimen3\font minus
  \fontdimen4\font\relax}
\providecommand{\BIBforeignlanguage}[2]{{%
\expandafter\ifx\csname l@#1\endcsname\relax
\typeout{** WARNING: IEEEtran.bst: No hyphenation pattern has been}%
\typeout{** loaded for the language `#1'. Using the pattern for}%
\typeout{** the default language instead.}%
\else
\language=\csname l@#1\endcsname
\fi
#2}}
\providecommand{\BIBdecl}{\relax}
\BIBdecl

\bibitem{preston2016}
``Resilience of the {{U}}.{{S}}. {{Electricity System}}: {{A Multi-Hazard
  Perspective}},''
  https://www.energy.gov/policy/downloads/resilience-us-electricity-system-multi-hazard-perspective.

\bibitem{hussain2019}
A.~Hussain, V.-H. Bui, and H.-M. Kim, ``Microgrids as a resilience resource and
  strategies used by microgrids for enhancing resilience,'' \emph{Applied
  Energy}, vol. 240, pp. 56--72, Apr. 2019.

\bibitem{hirsch2018}
A.~Hirsch, Y.~Parag, and J.~Guerrero, ``Microgrids: {{A}} review of
  technologies, key drivers, and outstanding issues,'' \emph{Renewable and
  Sustainable Energy Reviews}, vol.~90, pp. 402--411, Jul. 2018.

\bibitem{booth2017}
S.~Booth, ``Microgrid-{{Ready Solar PV}} - {{Planning}} for {{Resiliency}},''
  \emph{NREL/FS-7A40-70122}, p.~2, 2017.

\bibitem{DOER2020}
``Mobile {{Energy Storage Study}} | {{Mass}}.gov,''
  https://www.mass.gov/service-details/mobile-energy-storage-study.

\bibitem{abdeltawab2019}
H.~Abdeltawab and Y.~A.~I. Mohamed, ``Mobile energy storage sizing and
  allocation for multi-services in power distribution systems,'' \emph{IEEE
  Access}, vol.~7, pp. 176\,613--176\,623, 2019.

\bibitem{Kim_2019}
J.~Kim and Y.~Dvorkin, ``Enhancing distribution system resilience with mobile
  energy storage and microgrids,'' \emph{IEEE Transactions on Smart Grid},
  2019.

\bibitem{Che_2019}
L.~Che, M.~Shahidehpour, and M.~Shahidehpour, ``Adaptive formation of
  microgrids with mobile emergency resources for critical service restoration
  in extreme conditions,'' \emph{IEEE Transactions on Power Systems}, 2019.

\bibitem{Lei2018}
S.~Lei, J.~Wang, C.~Chen, and Y.~Hou, ``Mobile emergency generator
  pre-positioning and real-time allocation for resilient response to natural
  disasters,'' \emph{IEEE Transactions on Smart Grid}, 2018.

\bibitem{Lei2019}
S.~Lei, C.~Chen, H.~Zhou, and Y.~Hou, ``Routing and scheduling of mobile power
  sources for distribution system resilience enhancement,'' \emph{IEEE
  Transactions on Smart Grid}, 2019.

\bibitem{koutsoukis2019}
N.~C. Koutsoukis, P.~S. Georgilakis, and N.~D. Hatziargyriou, ``Service
  restoration of active distribution systems with increasing penetration of
  renewable distributed generation,'' \emph{IET Generation, Transmission \&
  Distribution}, vol.~13, no.~14, pp. 3177--3187, 2019.

\bibitem{choi2022}
M.-G. Choi, J.-H. Choi, S.-Y. Yun, and S.-J. Ahn, ``{{MILP-Based Service
  Restoration Method Utilizing Both Existing Infrastructure}} and {{DERs}} in
  {{Active Distribution Networks}},'' \emph{IEEE Access}, vol.~10, pp.
  36\,477--36\,489, 2022.

\bibitem{dugan2021}
J.~Dugan, S.~Mohagheghi, and B.~Kroposki, ``Application of {{Mobile Energy
  Storage}} for {{Enhancing Power Grid Resilience}}: {{A Review}},''
  \emph{Energies}, vol.~14, no.~20, p. 6476, Oct. 2021.

\bibitem{chen2016a}
C.~Chen, J.~Wang, F.~Qiu, and D.~Zhao, ``Resilient {{Distribution System}} by
  {{Microgrids Formation After Natural Disasters}},'' \emph{IEEE Transactions
  on Smart Grid}, vol.~7, no.~2, pp. 958--966, Mar. 2016.

\bibitem{meng2020}
S.~Meng, R.~{Roofegari nejad}, and W.~Sun, ``Robust {{Distribution System Load
  Restoration}} with {{Time-Dependent Cold Load Pickup}},'' \emph{IEEE
  Transactions on Power Systems}, pp. 1--1, 2020.

\bibitem{chen2017}
C.~Chen, J.~Wang, and D.~Ton, ``Modernizing {{Distribution System Restoration}}
  to {{Achieve Grid Resiliency Against Extreme Weather Events}}: {{An
  Integrated Solution}},'' \emph{Proceedings of the IEEE}, vol. 105, no.~7, pp.
  1267--1288, Jul. 2017.

\bibitem{yuan2016}
W.~Yuan, J.~Wang, F.~Qiu, C.~Chen, C.~Kang, and B.~Zeng, ``Robust
  {{Optimization-Based Resilient Distribution Network Planning Against Natural
  Disasters}},'' \emph{IEEE Transactions on Smart Grid}, vol.~7, no.~6, pp.
  2817--2826, Nov. 2016.

\bibitem{momen2021}
H.~Momen, A.~Abessi, S.~Jadid, M.~{Shafie-khah}, and J.~P.~S. Catal{\~a}o,
  ``Load restoration and energy management of a microgrid with distributed
  energy resources and electric vehicles participation under a two-stage
  stochastic framework,'' \emph{International Journal of Electrical Power \&
  Energy Systems}, vol. 133, p. 107320, Dec. 2021.

\bibitem{poudel2019}
S.~Poudel and A.~Dubey, ``Critical {{Load Restoration Using Distributed Energy
  Resources}} for {{Resilient Power Distribution System}},'' \emph{IEEE
  Transactions on Power Systems}, vol.~34, no.~1, pp. 52--63, Jan. 2019.

\bibitem{kleinberg2011}
M.~R. Kleinberg, K.~Miu, and H.-D. Chiang, ``Improving {{Service Restoration}}
  of {{Power Distribution Systems Through Load Curtailment}} of {{In-Service
  Customers}},'' \emph{IEEE Transactions on Power Systems}, vol.~26, no.~3, pp.
  1110--1117, Aug. 2011.

\bibitem{liu2021}
W.~Liu and F.~Ding, ``Collaborative {{Distribution System Restoration
  Planning}} and {{Real-Time Dispatch Considering Behind-the-Meter DERS}},''
  \emph{IEEE Transactions on Power Systems}, vol.~36, no.~4, pp. 3629--3644,
  Jul. 2021.

\bibitem{liu2021b}
------, ``Hierarchical {{Distribution System Adaptive Restoration With Diverse
  Distributed Energy Resources}},'' \emph{IEEE Transactions on Sustainable
  Energy}, vol.~12, no.~2, pp. 1347--1359, Apr. 2021.

\bibitem{glomb2022}
L.~Glomb, F.~Liers, and F.~R{\"o}sel, ``A rolling-horizon approach for
  multi-period optimization,'' \emph{European Journal of Operational Research},
  vol. 300, no.~1, pp. 189--206, Jul. 2022.

\bibitem{ahmadi2015}
H.~Ahmadi and J.~R. Mart{\i}{\textasciiacute}, ``Linear {{Current Flow
  Equations With Application}} to {{Distribution Systems Reconfiguration}},''
  \emph{IEEE Transactions on Power Systems}, vol.~30, no.~4, pp. 2073--2080,
  Jul. 2015.

\bibitem{li2021b}
Y.~Li, S.~Zhang, R.~Hu, and N.~Lu, ``A meta-learning based distribution system
  load forecasting model selection framework,'' \emph{Applied Energy}, vol.
  294, p. 116991, Jul. 2021.

\bibitem{erdener2022}
B.~C. Erdener, C.~Feng, K.~Doubleday, A.~Florita, and B.-M. Hodge, ``A review
  of behind-the-meter solar forecasting,'' \emph{Renewable and Sustainable
  Energy Reviews}, vol. 160, p. 112224, May 2022.

\end{thebibliography}
\bibliographystyle{IEEEtran}

\section*{Appendix}
\subsection{Proof of Claim-\ref{thm:SoC}}
\begin{proof}
The estimated SoC from stage-1 problem $\forall j \in \mathcal{N}^{ES}_{GFM}$ is given by (\ref{eqn:estSoC}) where, $SoC_{j,t}$ is the actual SoC of battery obtained from previous feedback from system,  $\hat{P}^D$ is the forecasted net load demand in the interval $[t, t+\Delta t]$.
\begin{multline}\label{eqn:estSoC}
    \hat{SoC}_{j,t+\Delta t} = SoC_{j, t} +\\  \frac{\hat{P}^D_t - \sum_{i\in\mathcal{N}^{DG}}P_{i,t}^{DG} - \sum_{i\in\mathcal{N}^{ES}_{GFL}}P_{i,t}^{ES}}{\overline{E}_j^{ES}\eta_j}\Delta t
\end{multline}

The actual measured SoC of battery $\forall j \in \mathcal{N}^{ES}_{GFM}$ can be formulated as shown in (\ref{eqn:measSoC}).
\begin{multline}\label{eqn:measSoC}
    SoC_{j,t+\Delta t} = SoC_{j, t} - \\ \frac{\int_{t}^{t+\Delta t}\left[P^{total}(t) - \sum_{i\in\mathcal{N}^{DG}}P_i^{DG}(t) - \sum_{i\in\mathcal{N}^{ES}_{GFL}}P_i^{ES}(t)\right]dt}{\overline{E}_j^{ES}\eta_j}
\end{multline}

Where, $P^{total}(t)$ is total net demand  in the system with power loss. Difference in SoC between measured and stage-1 estimation given by $\Delta SoC_{j,t+\Delta t}$ can be simplified as shown in eqn-\ref{eqn:SoCSimp}.
\begin{equation*}
  \Delta SoC_{j,t+\Delta t} = SoC_{t+\Delta t} - \hat{SoC}_{j,t+\Delta t}
\end{equation*}
\begin{multline}\label{eqn:SoCSimp}
  \Delta SoC_{j,t+\Delta t} = \left[ \frac{\hat{P}^D\Delta t - \int_t^{t+\Delta t}P^{total}(t)dt}{\overline{E}_j^{ES}\eta_j} \right]\\ - \left[\frac{ \sum_{i\in\mathcal{N}^{DG}\cup\mathcal{N}^{ES}_{GFL}} \left(\int_t^{t+1}P_i(t)dt - \hat{P}_{i,t}\Delta t \right)}{\overline{E}_j^{ES}\eta_j}\right]
\end{multline}

Now, $\int_t^{t+\Delta t}P^{total}(t)dt$ can be approximated by the average value $\overline{P}^{total}\Delta t$, where $\overline{P}^{total}$ is the average demand value in the interval $[t, t+\Delta t]$. Further, the grid-following resources dispatch in stage-2 over the horizon $\mathcal{K}$, $P^{G}_{i,k}$ is kept close to the scheduling value $\hat{P}^G_{i,t}$ $\forall i\in\mathcal{N}^G_{GFL}, k\in\mathcal{K}, t\in\mathcal{T}$ through stage-2 objective as shown in (\ref{eqn:stage2obj}). This ensures the stage-2 dispatch is close to scheduled value ($P_i(t) \approx \hat{P}_{i,t}$) which cancels out second term in (\ref{eqn:SoCSimp}).
\begin{equation}
\begin{aligned}
  \Delta SoC_{j,t+\Delta t}\overline{E}^{ES}_j\eta_j &\approx (\hat{P}^D - \overline{P}^{total})\Delta t - 0 \quad \text{since,} \quad P_i(t) \approx \hat{P}_{i,t}\\
  \text{let,  } \Delta P^D_t &\approx (\hat{P}^D - \overline{P}^{total})\\
  \Delta P^D_t &\approx  \frac{\Delta SoC_{j,t+1}\overline{E}^{ES}_j\eta_j}{\Delta t}\\
  \Delta P^D_t &\approx \kappa\Delta SoC_{j,t+1}
\end{aligned}
\end{equation}

Therefore, the average forecast and modeling error in the interval $[t, t+\Delta t]$ is approximately equal to the difference in SoC computed from feedback measurements and stage-1 results with a constant factor $\kappa = \frac{\overline{E}^{ES}_j\eta_j}{\Delta t}$.
\end{proof}

\end{document}